\documentclass[aip,jcp,amsmath,amssymb,12pt]{revtex4-1}

\usepackage{dcolumn}
\usepackage{bm}

\usepackage{graphicx}
\usepackage{amsfonts}
\usepackage{amsmath}
\usepackage{amssymb}
\usepackage[english]{babel}
\usepackage{graphicx, epsfig}
\usepackage{subfigure}
\usepackage{latexsym}
\usepackage{color}
\usepackage{epsfig}
\usepackage{amssymb}
\usepackage{graphics}

\newcommand{\mb}[1]{\mbox{\boldmath$#1$}}

\newcommand{\ds}{\displaystyle}
\newcommand{\beq}{\begin{eqnarray}}
\newcommand{\beqq}{\begin{eqnarray*}}
\newcommand{\eeq}{\end{eqnarray}}
\newcommand{\eeqq}{\end{eqnarray*}}

\newcommand{\x}{\mbox{\boldmath$x$}}

\newcommand{\X}{\mbox{\boldmath$X$}}
\newcommand{\w}{\mbox{\boldmath$w$}}

\def\ds#1{\displaystyle{#1}}

\begin{document}

\title{\bf Encounter dynamics of a small target by a polymer diffusing in a confined domain}
\author{A. Amitai$^{1*}$ C. Amoruso$^{1 *}$  A. Ziskind$^{2}$ D.
Holcman}
\affiliation{1:Ecole Normale Sup\'erieure, Institute of Biology (IBENS), Group of Computational Biology and Applied Mathematics, 46 rue d'Ulm, 75005 Paris, France. \\
2:Department of Neuroscience, Center for Theoretical Neuroscience, Columbia University, 1051 Riverside Drive Unit 87 Kolb Research Annex, New York, New York 10032, USA. \\
$^*$ These authors contributed equally.}


\begin{abstract}
We study the first passage time for a polymer, that we call the narrow encounter time (NETP), to reach a small target located on the surface of a microdomain. The polymer is modeled as a Freely Joint Chain (beads connected by springs with a resting non zero length) and we use Brownian simulations to study two cases:  when (i) any of the monomer or (ii) only one can be absorbed at the target window. Interestingly, we find that {in the first case} the NETP is an increasing function of the polymer length until a critical length, after which it decreases. Moreover, in the long polymer regime, we identified an exponential scaling law for the NETP as a function of the polymer length. {In the second case, the position of the absorbed monomer along the polymer chain strongly influences the NETP}. Our analysis can be applied to estimate the mean first time of a DNA fragment to a small target in the chromatin structure or for mRNA to find a small target.
\end{abstract}

\maketitle

\section{Introduction}
Polymers such as DNA or mRNA often have to move inside
constrained environment such as the nucleus or the cytoplasm before
reaching a small, strategic target. Such targets include nuclear
pores located on the nuclear envelope or ribosomes dispersed in the
cytoplasm involved in protein synthesis \cite{Bruce-Alberts}: mRNA must { exit } from the nucleus via passive diffusion in order to synthesize proteins (in the absence of any active transport of
the RNA in the nucleus or the cytoplasm \cite{sedat}). The task of finding a small nuclear pore can also
arise in the context of gene delivery, where DNA fragments have to enter the
nucleus. Finally, during the process of double strand DNA repair, { DNA ends have to search for one another in the confined chromatin environment \cite{almouzni}  and can also localize to the membrane periphery to interact with the nuclear pores \cite{KarineDUBRANA}}.\\
Previous studies of polymer in confined domains focused on their static properties \cite{Raphael,Dayantis1985}, translocation \cite{KantorKardar,Muthukumar,PRE1998}, reptation \cite{deGennes_book} through a cylindrical tube and their dynamics \cite{Spakowitz2010}. During translocation, the polymer is threaded through a pore and diffuses until it exits the other side. The polymer has to overcome an effective barrier generated by each monomer on the surface membrane. In contrast, the encounter process we are studying here differs significantly and we shall study the search process of a small target by a freely diffusing polymer. This search process shares some similarities with the polymer looping problem \cite{Pastor1996,Toan,Amitai2012} where the critical time scale is defined by the duration for the two polymer ends to meet.
We present here a numerical study for the motion of a Freely Joint Chain \cite{Toan} polymer in a confined microdomain. Two different types of kinetics for diffusion-controlled processes in dense polymer systems have been discussed in \cite{deGennes82}, depending on the root mean square displacement of the active polymer site (compact and non compact exploration). When the polymer is very small and its motion is dominated by its center of mass diffusion, then the narrow encounter time {of the polymer } (NETP) is precisely the mean first passage time for a Brownian particle to a small target, also known as the narrow escape time (NET \cite{SSH1,SSH2,SSH3,PNAS,HS,Ward1,Pillay,Grigoriev,Reingruberabad}. In a space of dimension $d$, it is given by
\beq
 \label{NET_d2}
 \langle \tau_{2d}\rangle &=&\ds{\frac{A}{\pi D}\ln \frac{1}{\varepsilon }+\mathcal O(1)  } \mbox{ for } d=2
 \\ \nonumber\\
 \langle \tau_{3d}\rangle &=& \ds{\frac{|\Omega|}{4\varepsilon D \left[1+\ds{\frac{L(\mb{0})+ N(\mb{0})}{2\pi}} \,\varepsilon \log \varepsilon+\mathcal O(1) \right]}} \mbox{ for } d=3,
 \label{NET_d3}
 \eeq
where $A$ is the area (in two dimensions) and $\varepsilon$ is the
ratio of the absorbing region of the boundary to the total length of
the boundary. In three dimensions, $|\Omega|$ is the volume, $\varepsilon$ is
the radius of the small absorbing  target and $L({\bf 0})$ and $N({\bf
0})$ are the two principal curvatures at the origin. $D$ is the
diffusion constant. The order one term in each expression depends on
the initial position of the moving particle \cite{Pillay}. These
formulas have been extended to the case of multiple windows \cite{Pillay,PLA-holes,JPA-holes}.
We study here how the polymer length controls the mean time for a monomer to reach a small target (a  disk of radius $\varepsilon$) located on the boundary of a microdomain. Once a monomer hits this disk, the target is found. In the context of chemical reactions, this time is the reciprocal of the forward reaction rate for diffusion limited processes. The present analysis of the NETP complements previous computations of the
forward rate \cite{Wilemski, Zwanzig, Zwanzig2,Szabo, Szabo2,Perico}. It may also be used to estimate the first time to activate a gene by a factor located on the DNA, which is different from the classical activation due to a transcription factor \cite{Malherbe,Berg,Mirny,Benichou}. However, at this stage a final analytical formula has to be derived \cite{KupkaHolcmanSmale}.

The paper is organized as followed: we first present the Freely Joint Chain polymer, where the monomers are represented by beads connected by elastic springs. We run Brownian simulations and estimate the NETP in two cases:
\begin{enumerate}
  \item When \textit{any} of the polymer monomer can be  absorbed at the
  target region, which we designate $\langle \tau_{\textrm{any}}\rangle$.
  \item When only one of the monomers can escape through the
  target region, which we call $\langle \tau_{\textrm{mon}} \rangle$.
\end{enumerate}
Each of these cases might be relevant under different conditions: in some cases one of the two ends of the polymer has to reach the target, whereas in other situations, it might be sufficient for any part of the
polymer to reach the target, such as in gene activation. We then present the NETP simulation results. To study the effect associated to the polymer length, we estimate the mean first passage time of a single absorbing monomer to the boundary of a total absorbing sphere, when the other monomers are reflected. We present various empirical results about the motion of a single monomer. Furthermore, we study the dependence of the NETP on the monomer's position along the chain and the distribution of the arrival times. Finally, we extend the NETP simulation analysis to include bending elasticity.

\begin{figure}
	\centering
		\includegraphics[scale=0.7]{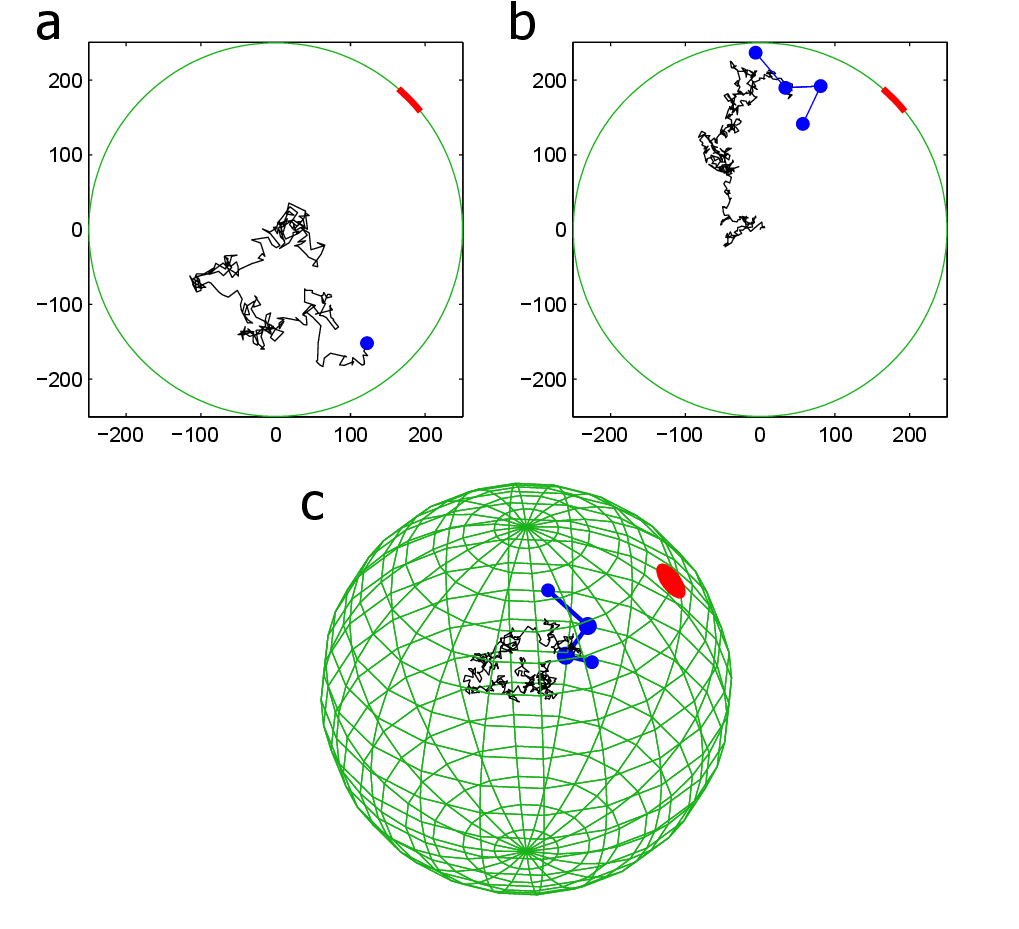}
	\caption{{\bf Confined trajectories of FJC polymer. }
 In a circular(2D)/spherical(3D) domain of radius $250$nm, the target size is $\varepsilon = 50$nm. Trajectories of the center of mass (black) before absorption at the target (red). (a) Simulation with 1 bead,
moving in 2 dimensions. (b-c) Simulations of a 4-bead FJC polymer,
moving in 2 dimensions (b) and in 3 dimensions (c).}
	\label{fig_snapshots}
\end{figure}

\section{Results}

\subsection*{The Narrow Encounter Time (NETP) for different polymer length}
To study the NETP, we model the polymer motion with a Freely Joint Chain (FJC) \cite{Toan}
inside a convex domain $\Omega$. 
Each polymer consists of $N$ beads with coordinates $\x=(\x_1 ,..\x_N )$, where consecutive beads are
connected by springs of characteristic constant $k$ and
experience isotropic random collision forces. These forces are
modeled as Brownian motions. The average bond length between neighboring monomers is given by
\beq
\langle \vert \x_k - \x_{k+1} \vert \rangle \, =l_0.
\eeq
Because the DNA molecule is a quite unextendable at small scales below the persistence length of 50nm, we decided to include this length in our simulations. Thus, contrary to the usual Rouse model \cite{Doi:Book} in which $l_0 = 0$, we study here the FJC where the equilibrium length between the nearest neighbors is nonzero with $l_0 = 50$nm. For long polymer chains, it should not matter much which of the models (FJC or Rouse) to use, because they share many similarities \cite{Doi:Book}.

The dynamics of the polymer is described by the over-damped Langevin-Smoluchowski equation:
\beq
\dot{\x} +\frac{1}{\gamma}\nabla U^N=\sqrt {2D} {\rm {\bf \dot { \boldsymbol{\eta}}}},
\label{BD.eq}
\eeq
where ${\bf \eta}$ is a Gaussian random variable with zero mean and unit
variance, $\gamma$ is the friction coefficient and $U^{N}$ is the potential energy generated by the springs
\beq
\ds{ U^N(\x)=\frac{k}{2}\sum_{k=1}^{N-1} \left(\vert \x_k-\x_{k+1} \vert-l_0 \right)^2 }.
\label{elastic}
\eeq
For all simulations, we use a circular or spherical microdomain $\Omega$.  To construct the initial configuration of the polymer, we put all monomers close to the origin folded on the top of each other and we then ran simulations for a time long enough that depend on the size of the polymer. Indeed, for short polymers, we run simulations during a period of time  $\frac{R^2}{D_{CM}}=NR^2/D$ ($R$ is the radius of the sphere), which represent the mean time for the center of mass to explore the spherical domain. For longer polymers for which the radius of gyration is comparable to $R$, we run a preliminary simulation for a time equal to the longest relaxation time of the polymer $\frac{l^2_0 N^2}{D}$ \cite{Doi:Book}. To conclude, to make sure that the initial configuration of the polymer inside the sphere was chosen at equilibrium, we run an initial simulation up to time $\textrm{max}\{\frac{NR^2}{D},\frac{l^2_0 N^2}{D}\}$.

Most of the boundary is reflecting, except for the small absorbing window, where the polymer can be absorbed. Sample trajectories (Fig. 1) of the center of mass
(initial position at the center of the domain) for different sets of parameters with beads and springs (blue), the boundary (green) and
the absorbing target (red). The analytical estimation of the NETP remains a challenging problem, since the computation involves generalizing the narrow escape theory to the tubular neighborhood of the absorbing hole in dimension $2N$ or $3N$ \cite{KupkaHolcmanSmale}. However, this computation does not fall into the narrow escape methodology \cite{SSH1,SSH2,SSH3,PNAS,HS,Ward1,Pillay,Grigoriev,Reingruberabad}, rather the computation requires to analyze the mean first passage time of a stochastic particle in a narrow band in high dimension, and thus many of the approximations used for a punctual particle are no longer valid. Obtaining such a formula for the polymer would be useful to better explore the complexity of the parameter space.

\subsection*{The NETP has a bell shape profile}
To study the dependency of the NETP as a function of the polymer
length $N$, we use the Euler's scheme (equation \ref{eqnum}) and run
Brownian simulations for a range of polymer lengths from $N = 1$ to
$N= 350$, in dimensions two and three under the two scenarios described above. In Figs. \ref{any.fig}-\ref{SingleMonDyn}, the NETP values are normalized to the NET
for a single particle, which we obtained from Brownian simulations: $\tau_{0} = 3 \mathrm{s}$ in
dimension two and $\tau_{0} = 15 \mathrm{s}$ in dimension 3, with
parameters listed in table 1 and the diffusion constant for a single monomer is $D=0.04 \mathrm{\mu m}^2/\mathrm{s}$ \cite{DNAdiffusionconst}. The NETP curve contains two phases that we shall now  discuss (the NET $\tau_{0}$ for one particle is referred in the figures as $\langle\tau_{2d}\rangle$ and $\langle\tau_{3d}\rangle$ in dimension 2 and 3 respectively).
\subsubsection*{The initial phase of the NETP increases linearly with the polymer length}
In dimensions two and three, the NETP is initially an increasing function of the number of beads $N$, until a critical value $N_c$ after which
it is decreasing. With the parameter of table 1, using Brownian simulations, we found that the critical value
$N_c \in [10-15]$ (later on, we obtained the same critical length when only a single monomer can be absorbed). We were surprised to see that the polymer size associated with this critical number is much smaller than the critical length { obtained from the radius of gyration $R_\textrm{g} \approx \sqrt{N_c}l_0/\sqrt{6}\approx 150$}. This suggests that the radius of gyration is not sufficient to characterize the effect of the confinement on the polymer dynamics.

In addition, for short polymer lengths, such that $N< N_c$, the NETP is largely determined
by the motion of the center of mass. When the polymer is far from the absorbing target,
none of the beads will be able to reach it, until the center of mass
has moved close to the target. The NETP thus reflects the mean first
passage time of the center of mass to the target. In this limiting
case, the center of mass undergoes Brownian motion with a diffusion
constant inversely proportional to the number of beads: $D_{\textrm{CM}}
=\frac{D}{N}$. Thus, in the regime $N\ll N_c$, the NETP is
approximately that of a single particle, but with a smaller
diffusion constant. The expression for the {NET} (\eqref{NET_d2} and
\eqref{NET_d3}) are inversely proportional to $D_{\textrm{CM}}$, and thus we
obtain the initial linear regime in $N$:\beq
\langle\tau_{\textrm{any}}\rangle_{2d} \approx N \langle\tau_{2d}\rangle\\ \nonumber
\langle\tau_{\textrm{any}}\rangle_{3d} \approx N \langle\tau_{3d}\rangle
\eeq
as confirmed in Fig. \ref{any.fig}. However for a polymer of length comparable to the size of the microdomain, the location of the center of mass does not determine anymore the NETP. In that regime, smaller subsections of the polymer
can be close to the target even if the center of mass is far away.
In addition, when any bead can be absorbed, increasing the polymer
length results in a decrease in the NETP (Fig. \ref{any.fig}). Interestingly, two regimes can be further distinguished for the decay phase.

\subsubsection*{The decay phase of the NETP is approximated by an exponential}
\begin{figure}[htbp]
\begin{center}
       {\includegraphics[scale=0.4]{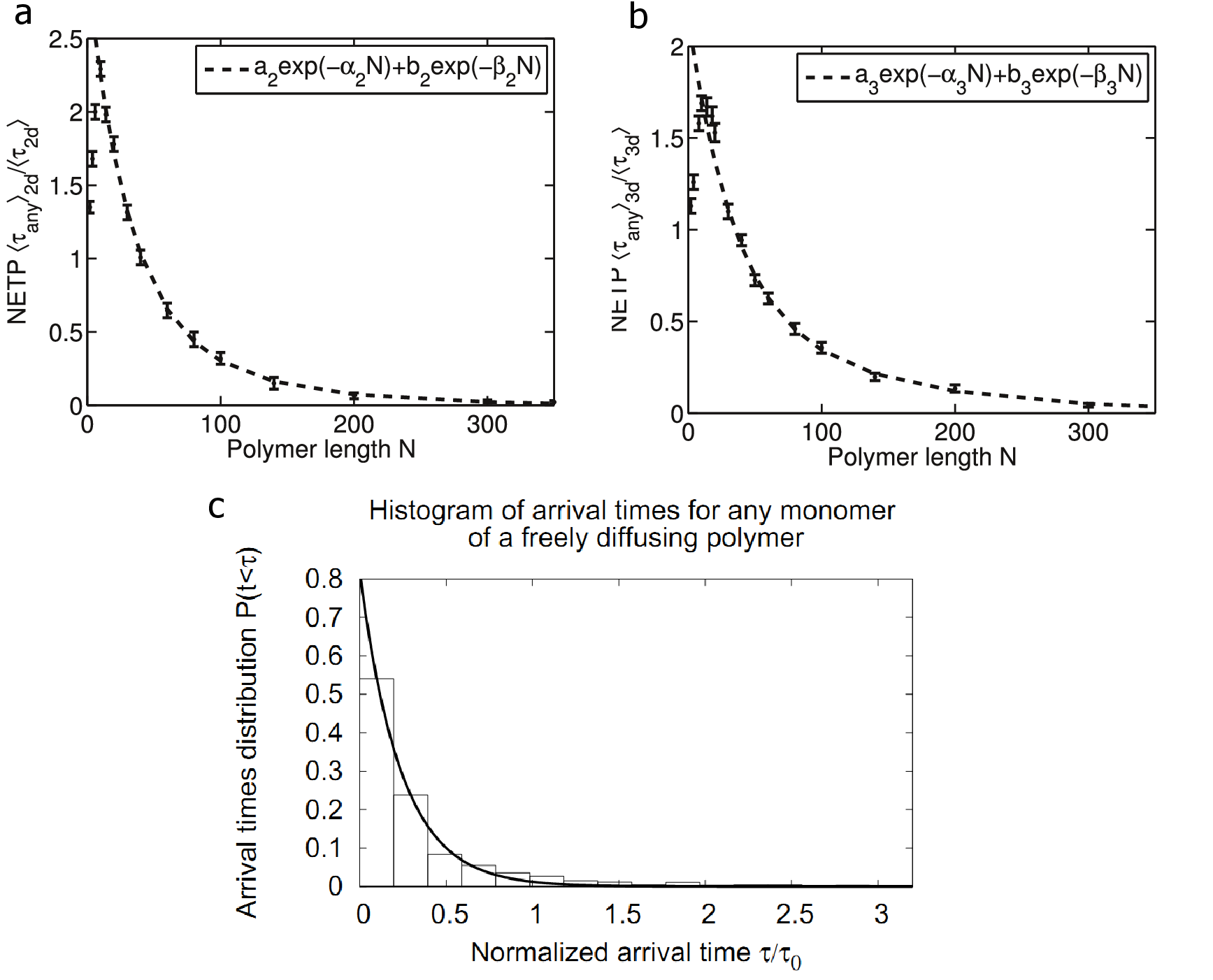}}
       \caption{\textbf{NETP for any monomer to the target, for different polymer lengths }
The results are normalized to $\tau_0$ (the NET for 1 bead), equations \eqref{NET_d2} and \eqref{NET_d3}. Depicted is the mean time
for any of the beads to reach the target in dimension two {\bf (a)} and
three {\bf (b)}. Each data point represents the average over 2000 runs.
The Brownian data is fitted by a double exponential function in
dimension 2, $\frac{\langle \tau_{\textrm{any}}(N)\rangle_{2d}}{\langle \tau_{2d}\rangle} = a_2\exp(-\alpha_2 N) +
b_2\exp(-\beta_2 N)$ with exponents $\alpha_2 = 0.0075, \beta_2 =
0.024$ and coefficients $a_2 = 0.23, b_2 = 2.17$ and in dimension 3
$\frac{\langle \tau_{\textrm{any}}(N)\rangle_{3d}}{\langle \tau_{3d}\rangle} = a_3\exp(-\alpha_3 N) + b_3\exp(-\beta_3
N)$ with exponents  $\alpha_3 = 0.0082, \beta_3 = 0.030$ and
coefficient  $ a_3= 0.60, b_3 = 1.56$.
{\bf (c)} Probability distribution $P[\tau/\tau_0]$ of arrival times for any monomer to a small target (in three dimensions) for $N=100$. The Probability distribution of the arrival times to a small hole
located on the boundary of a sphere in three dimensions, is approximated by a sum of two exponentials. The data is well approximated by
a single exponential of the form $Pr\{ \tau_{3d}=t\} = a \exp(-\alpha t)$ with $a =    0.814$, and $\alpha =   4.185$.}
\label{any.fig}
\end{center}
\end{figure}

\begin{figure}
	\centering
		\includegraphics[scale=0.5]{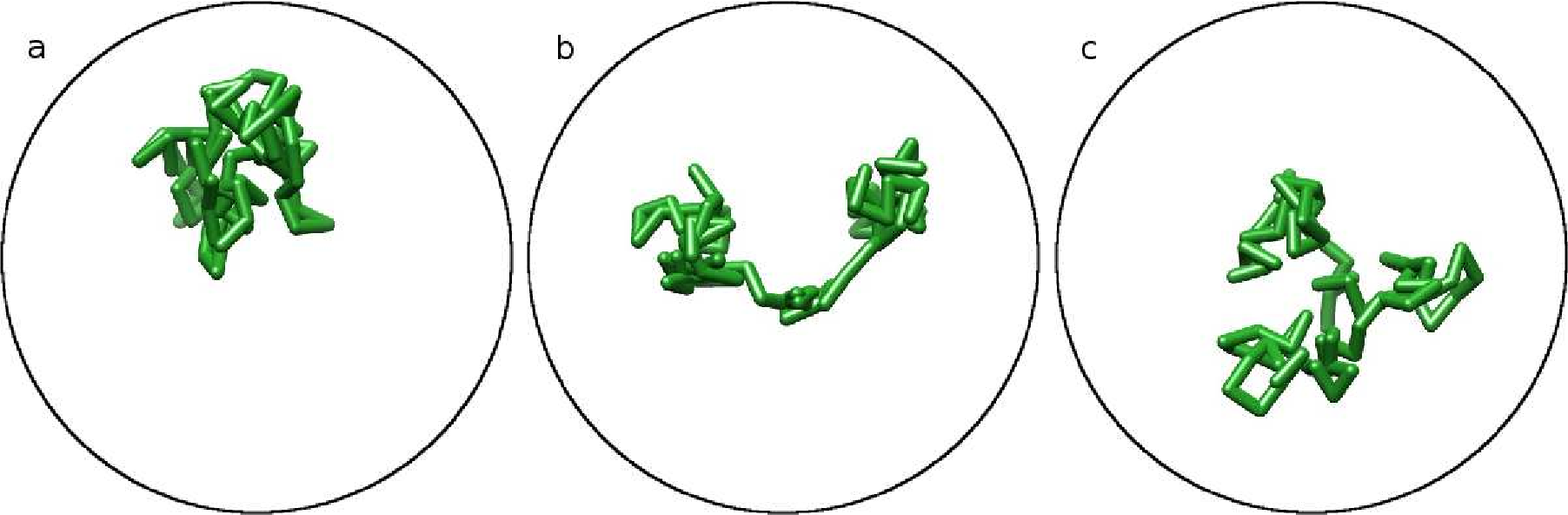}
	\caption{\textbf{Snapshots of the polymer configuration during the Brownian dynamics.}
A polymer ($N=100$) is shown at {\bf (a)} 7.5ms , {\bf (b)} 30ms  and  {\bf (c)} 60ms  during the Brownian dynamics (the initial configuration of the polymer has uniformly distributed angles, with length $l_0$). This simulation reveals Pearl-Like Structures (PLSs): a large PLS in (a), two PLSs (b) and three small ones (c). The time step is $\Delta=1.5 \times 10^{-4}$s.}
	\label{pearl.fig}
\end{figure}

{When any bead can be absorbed}, the decay phase of the NEPT (Fig. \ref{any.fig}) can be separated
into two different regimes that can be described as followed.  In the first one, the polymer moves
freely until a monomer hits the absorbing boundary. The NETP is determined by the competitive effects of a
decreased diffusion constant for the center of mass and an increased total polymer length. Increasing the polymer length leads to an effective smaller volume of the effective confining domain in which the
polymer has to find the absorbing window. To investigate more specifically the dynamics of the polymer, we
monitored in our simulations the time evolution of several polymers and observed at equilibrium various heterogeneous configurations with multiple Pearl-Like-Structure (PLS) (Fig. \ref{pearl.fig}). Thus, a polymer forms transient PLSs, sometimes a single PLS (Fig \ref{pearl.fig}a), or two transient small ones (Fig \ref{pearl.fig}b) or even three little ones (Fig \ref{pearl.fig}c). We conclude that a polymer of intermediate size is not simply uniformly distributed, rather its shape changes, making transient between various
substructures, characterized as PLSs These transients structures  (Fig \ref{pearl.fig}c) suggest that the NETP as a function of $N$ for a confined polymer cannot result from a simple scaling law, that can be analyzed from a steady state distribution.

In a second regime, when the length of the polymer becomes long enough, so that at least one monomer can always be found in the boundary layer (of size $\varepsilon$) of the absorbing hole \cite{SSH2,SSH3},  we expect the NETP to have  a different decay as a function of $N$ compared to the previous intermediate regime. In that case, the center of mass is strongly restricted due to the interaction of all the monomers with the microdomain surface (Fig. \ref{SingleMonDyn}d). The NETP is determined by mean time for a monomer in the boundary layer of the absorbing window. It is still an unsolved problem to obtain an asymptotic estimate for that time.
Using an optimal fit procedure, we obtain for the NETP  an empirical scaling approximation with two exponentials
\beq
\frac{\langle \tau_{\textrm{any}}(N)\rangle_{2d}}{\langle \tau_{2d}\rangle}= \ds{ a_2 \exp(- \alpha_2 N) + b_2 \exp(- \beta_2 N)} \hbox{ for }  \, d=2 \label{emp1}
\eeq
and
\beq
\frac{\langle \tau_{\textrm{any}}(N)\rangle_{3d}}{\langle \tau_{3d}\rangle}= \ds{ a_3 \exp(- \alpha_3 N) + b_3 \exp(- \beta_3 N)}  \hbox{ for } \, d=3. \label{emp2}
\eeq
${\langle \tau_{3d}\rangle}$ and ${\langle \tau_{2d}\rangle}$ are the NET for a single particle  $\tau_{0}$, in dimension 3 and 2 respectively. In dimension two, the exponents are $\alpha_2 = 0.0075, \beta_2 = 0.024$ and coefficients $a_2 = 0.23, b_2 = 2.17$ and in dimension
3, $\alpha_3 = 0.0082, \beta_3 = 0.030$ and coefficient  $ a_3 = 0.60, b_3 = 1.56$ (Fig. \ref{any.fig}).
The empirical laws \eqref{emp1}-\eqref{emp2} remains to be derived analytically.

\subsubsection*{ The arrival time is approximately Poissonian when any monomer can reach the target}
Finally, we have shown in Fig. \ref{any.fig}c, that histogram of arrival time of any monomer to the small target can well approximated by a single exponential, suggesting that the arrival time be almost Poissonian. This result can appear quite surprising, but it is the effect of the small target, which select the long time behavior of the polymer \cite{PNAS}.

\subsubsection*{Phenomenological explanation of NETP bell shape}
The bell shape nature of the NETP can be qualitatively explained
using the NET equations (\ref{NET_d2} and \ref{NET_d3}). {Indeed, a small polymer can be considered as a quasi-particle of radius $R_g = l_0\sqrt{N/6}$ \cite{deGennes_book}, evolving in an effective domain which is the full domain  minus its volume.  Thus the NETP is related to the mean first passage time of the {quasi-}particle with diffusion constant  $D_{\textrm{CM}} = \frac{D}{N}$ in the apparent domain of volume $V_a=\frac{4\pi}{3}(R-R_g)^3$, leading to a mean time proportional
\beq\label{volume_quasi}
 \frac{V_a}{\varepsilon D_N}=N\frac{4\pi}{3\varepsilon D}(R-l_0\sqrt{N/6})^3.
\eeq
This phenomenological formula shows that the mean time has a maximum for $N_m=25$ (parameter of table 1), which is an over estimation of the empirical value that we obtained from Brownian simulation $N=10$.  To summarize, when each bead can reach the absorbing target, the NETP exhibits two different behaviors: while for short polymer size the dynamics can be abstractly described by a quasi-particle, leading to a maximum of the NETP, for longer polymers the NETP decays with a single exponential, a behavior that was not expected. An analytical derivation of this result is still missing. }

\subsection*{When a single monomer can be absorbed, the NETP depends on its location on the polymer chain}
\begin{figure}[htbp]
\begin{center}
       {\includegraphics[scale=0.35]{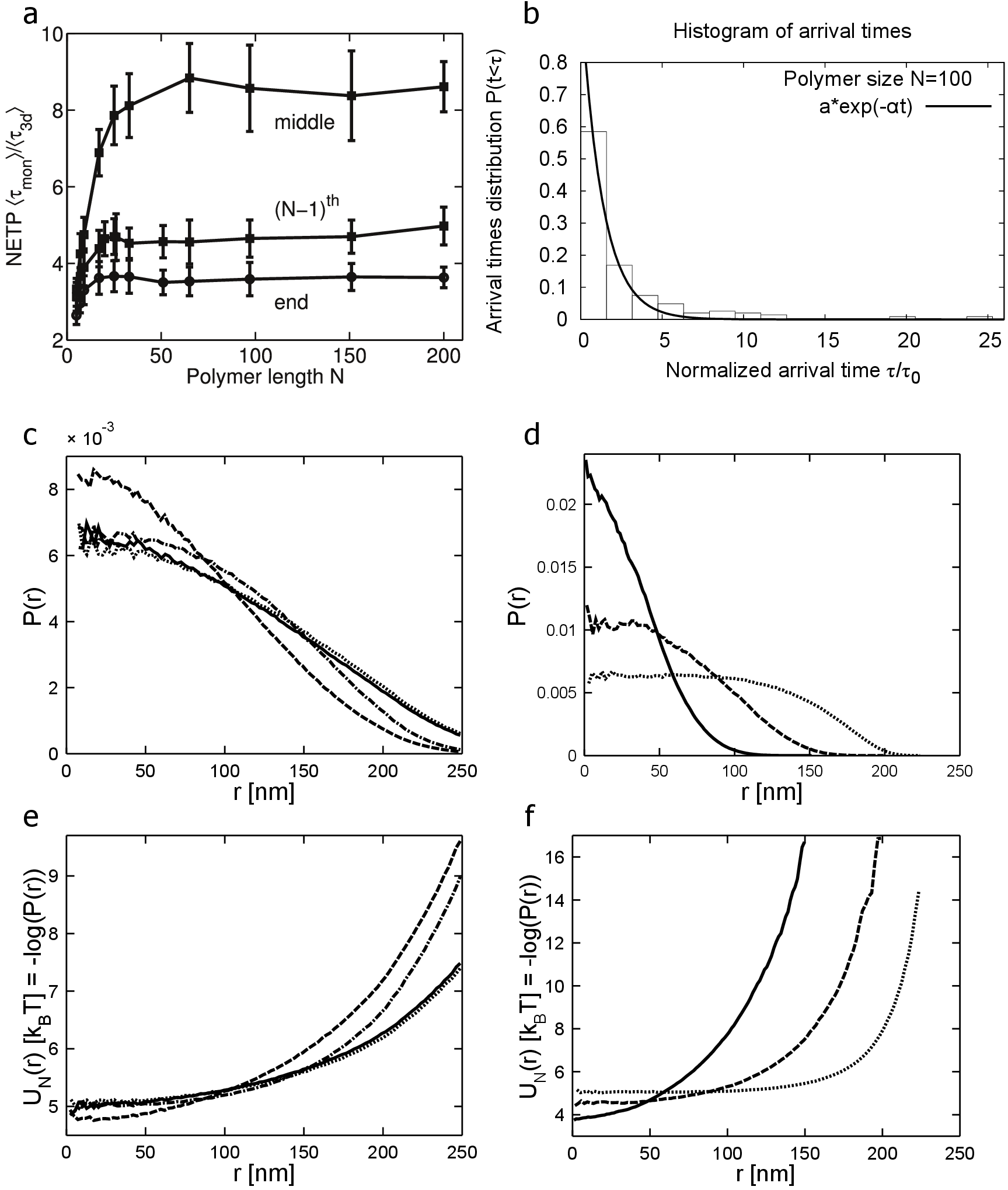}}
\caption{\textbf{Dynamics and distribution of a single monomer}:  \textbf{(a)} NETP for the three monomers: end, middle and $N - 1$ as a function of the polymer length $N$ in three dimensions (Brownian simulations): The encounter time is normalized to the time $\tau_0$ (for a single bead). Parameters are described in table 1. {\bf (b)} Probability distribution $P[\tau/\tau_0]$ of arrival times for the end monomer to a small target (in three dimensions).  The data is well approximated by a single exponential of the form $Pr\{ \tau_{3d}=t\} = a \exp(-\alpha t)$ with $a = 1.014, \alpha = 0.76$.
\textbf{Monomers radial distribution}: {\bf (c)} Probability distribution function (pdf) of monomers position. The radial pdf is calculated from Brownian simulations for the end monomers $N=16$ (points), $N=150$ (full line) and the middle monomers $N=16$ (points-line), $N=150$ (dashed line). {\bf(d)} The pdf of the center of mass for $N=16$ (points), $N=48$ (dashed line), $N=150$ (full line). The normalized effective potential $U_N(r)=-k_B T \log(P(r))$ acting on a monomer was calculated from the radial pdf. {\bf(e)} The potential for the end and middle monomers corresponding to {\bf(c)}. {\bf(f)} The potential for the center of mass corresponds to {\bf(d)}. The asymptotic behavior for the end and the middle monomers are significantly different, confirming that the boundary has different effect depending on the position of each monomer.}
\label{SingleMonDyn}
\end{center}
\end{figure}

To study the impact on the NETP of the monomer location along the polymer chain that can be absorbed at the target, we ran a new set of Brownian simulations (Fig. \ref{SingleMonDyn}). The polymer is confined in a sphere and only one monomer (we call it active monomer) can be absorbed, while all others are reflected at the target site. Interestingly, the simulation reveals that the location of the absorbing monomer on the chain can drastically affect the arrival time to the small target. Between the middle and the end monomer, we observe a factor 3 reduction in the arrival time. Interestingly, already taking the $N-1$th monomer compared to the last one is making a noticeable difference. In addition, the NETP increases to a different plateau that depends also on the monomer location: the plateau starts at a length $N=20$ for the end monomer, while it is around $N=75$ for the middle monomer. We note that the effect of the boundary starts on the polymer dynamics is seen already for $N=20$, which is much smaller than the length of a polymer for which the gyration radius is comparable to the domain radius.

To analyze this behavior, we examined each phase separately.  In the increasing phase (as a function of the polymer length, Fig. \ref{SingleMonDyn}a), the arrival of the monomer to the target is mainly governed by the diffusion of the polymer center of mass, which can be approximated by a diffusing ball of diffusion constant $D_{\textrm{CM}} = D/N$. However, for larger $N$, this approximation is not valid, rather the arrival time converges to a constant value that depends on the structure of the polymer where the stochastic dynamics is much richer than simple diffusion.  The motion of the active monomer, which is affected by the entire polymer depends on its length and the boundary of the confined domain \cite{Amitai2010}. Intuitively, the interaction of the active monomer with the other monomers generates an effective potential through their interaction with the boundary, which is different for the middle and the end monomer, leading to the major differences reported in Fig. \ref{SingleMonDyn}a and c.

For $N$ large enough, the middle monomer is more confined than the end one, and by analogy with the diffusion of a stochastic particle in a potential well, the middle monomer has to surmount a higher potential barrier to reach the small target located on the boundary. To clarify this hand waving explanation, we approximate the active monomer motion as diffusion in a spherical symmetrical potential $V$ inside our confined spherical domain for the polymer. The stochastic description is
 \beq
 \dot{\X} =\frac{V(\X)}{\gamma }+\sqrt{\frac{2kT}{\gamma}} \dot{\w},
 \eeq
 where $\w$ is the standard Brownian motion. Using the symmetry of the domain, the potential $V$ has a single minimum at the center. Interestingly, in the high potential barrier approximation, the mean time to a small target does not depend on the specific shape of the potential, but rather on its minimum and maximum \cite{SingerSchuss}. In that case, the mean time $\tau (N)$ to a target, which depends on the polymer size is given by:
 \beq \label{NETactivation}
\tau(N)= \frac{(2\pi)^{3/2} \gamma \sqrt{kT}}{4a \omega_N^{3/2}}\exp\left[ \frac{U_N(r)}{kT}\right]
\eeq
where $U_N(r)$ is the energy barrier generated by the polymer due to the presence of the boundary of the domain at the target site and $\omega_N$ is the frequency at the minimum ($U_N(r)\approx\frac{\omega_N}{2}r^2$ near 0) \cite{SingerSchuss}.  The potential $U_N(r)$ can be recovered from formula \ref{NETactivation} and the numerical simulations described in Fig. \ref{SingleMonDyn}c.  We conclude that the position along the polymer chain of the monomer interacting with the target critically influences the search process.

We now directly evaluate the monomers distribution inside our confined domain using a numerical approach. To further characterize the dynamics of a single monomer, we computed the equilibrium radial probability distribution function (pdf) $P(r)$ of the monomers position. We found that the monomers, depending on their location along the chain, are differently distributed inside the confined domain (Fig.\ref{SingleMonDyn}c,d). The middle monomer is more restricted to the center, away from the boundary compared to the end monomer, which explores in average a larger area. As the length of the chain increases, the middle monomer gets more restricted, while the end one does not seem to be much affected (Fig.\ref{SingleMonDyn}c). The center of mass has a similar behavior as the middle monomer and its spatial distribution gets restricted for increasing polymer length. From the radial pdf, we decided to evaluate the effective potential $U_N(r)$ acting on a monomer by $U_N(r) = -k_BT\log(P(r))$ defined in equation \ref{NETactivation} The potential acting on the center of mass is large enough so that it never arrived to the periphery during our simulations (Fig.\ref{SingleMonDyn}e,f).
Finally, similar to the case where all the monomers can interact with the target, we show in Fig.\ref{SingleMonDyn}b, that the arrival time distribution is well approximated by a single exponential, showing that the distribution is almost Poissonian. However the rate depends on the location of absorbing monomer along the polymer chain.

\begin{figure}[htbp]
\begin{center}
       {\includegraphics[scale=0.55]{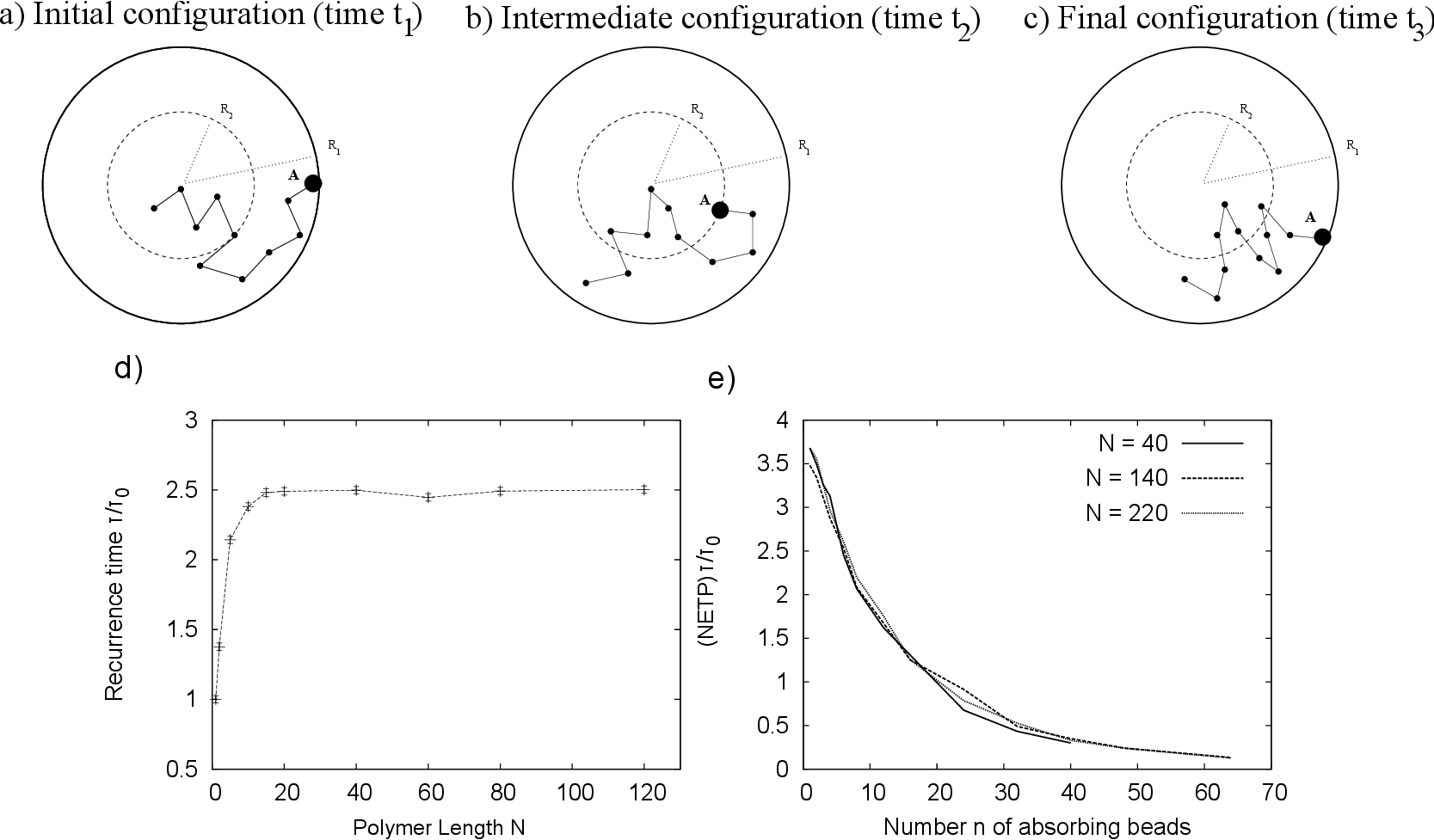}}
\caption{{\bf The recurrence time of a polymer between two concentric disks and the influence of the absorbing polymer size. } A perfectly flexible polymer is introduced into the spherical domain and we preform a preliminary Brownian simulation until the equilibrium regime is achieved. After that, we estimate the first time $t_1$ that one of the
end monomer hits the surface of the external sphere {\bf (a)} then the first time $t_2$ that the it enters into the sphere of
radius $R_2$ {\bf (b)} and finally the first time $t_3$
that it hits again the boundary of the external sphere {\bf (c)}. The recurrent
time is average of $t_3 - t_1$ over many realizations. {\bf (d)} recurrence time for the
end monomers (polymer of length N = 60) in dimension three, computed as
described in panel a-c.  {\bf (e)} Effect of the polymer reacting site size on the arrival time to a fully absorbing three dimensional sphere.
The polymer is composed of $N$ monomers, with $n$ absorbing ones, the other
$N-n$ are reflected at the external sphere. We present the NETP to the boundary for three polymer sizes (40, 140 and 220). The results are normalized by the NET for a single Brownian particle $\langle\tau_{3d} \rangle$.}
\label{rec_time}
\end{center}
\end{figure}

\subsubsection*{Influence of the polymer length on the arrival time of an ensemble of absorbing monomers to the sphere boundary}
To evaluate the effect of the polymer length on the monomer dynamics, we study the arrival time of the ensemble of $n$ consecutive (the first one a polymer end) monomers to a sphere.  As described in the previous section, the reflecting monomers can prevent the absorbing ones to reach the boundary of the sphere. To explore the arrival time as a function of $n$, we run Brownian simulations in a three dimensional sphere for various polymer sizes ($N$ monomers) and absorbing lengths $n$, while the  $N-n$ remaining monomers are reflected on the surface.  We have found (Fig.\ref{rec_time}a) that the arrival time decays with $n$, independently of the polymer size ($N=40, 140, 220$).

To further investigate the role of the polymer length on the arrival time, we studied the recurrence time of a single monomer, which is the time for that monomer to move from a sphere of radius $R_2=\frac{R_1}{2}$ to the boundary (radius $R_1=250$nm), as described in  Fig.\ref{rec_time}b-d. The recurrence time is increasing with the size of the polymer. However, it becomes independent on the polymer length for $N\geq 20$ (Fig.\ref{rec_time}e)  while the motion of a single monomer does not depend on the size of polymer. In the regime of large $N$, the longest relaxation time due to the internal modes of the polymer \cite{Doi:Book} is larger than the time it takes for a monomer to be absorbed at boundary (this situation happens when the radius of gyration of the polymer is larger than the radius of the sphere). To clarify that the time to absorption is independent of the total polymer length $N$, we simply recall that  the motion of a single monomer in the large N regime is that of a correlated particle, that can be described as anomalous \cite{Amitai2010}. In that case,  the Mean Square displacement is shown  \cite{Amitai2010} to be independent of the polymer length. To conclude, Fig.\ref{rec_time}d shows that the reflecting monomers prevent the absorbing ones to reach the surface of the sphere only below a certain number.

\subsection*{NETP with additional stiffness}
\begin{figure}[htbp]
\begin{center}
{\includegraphics[scale=0.50]{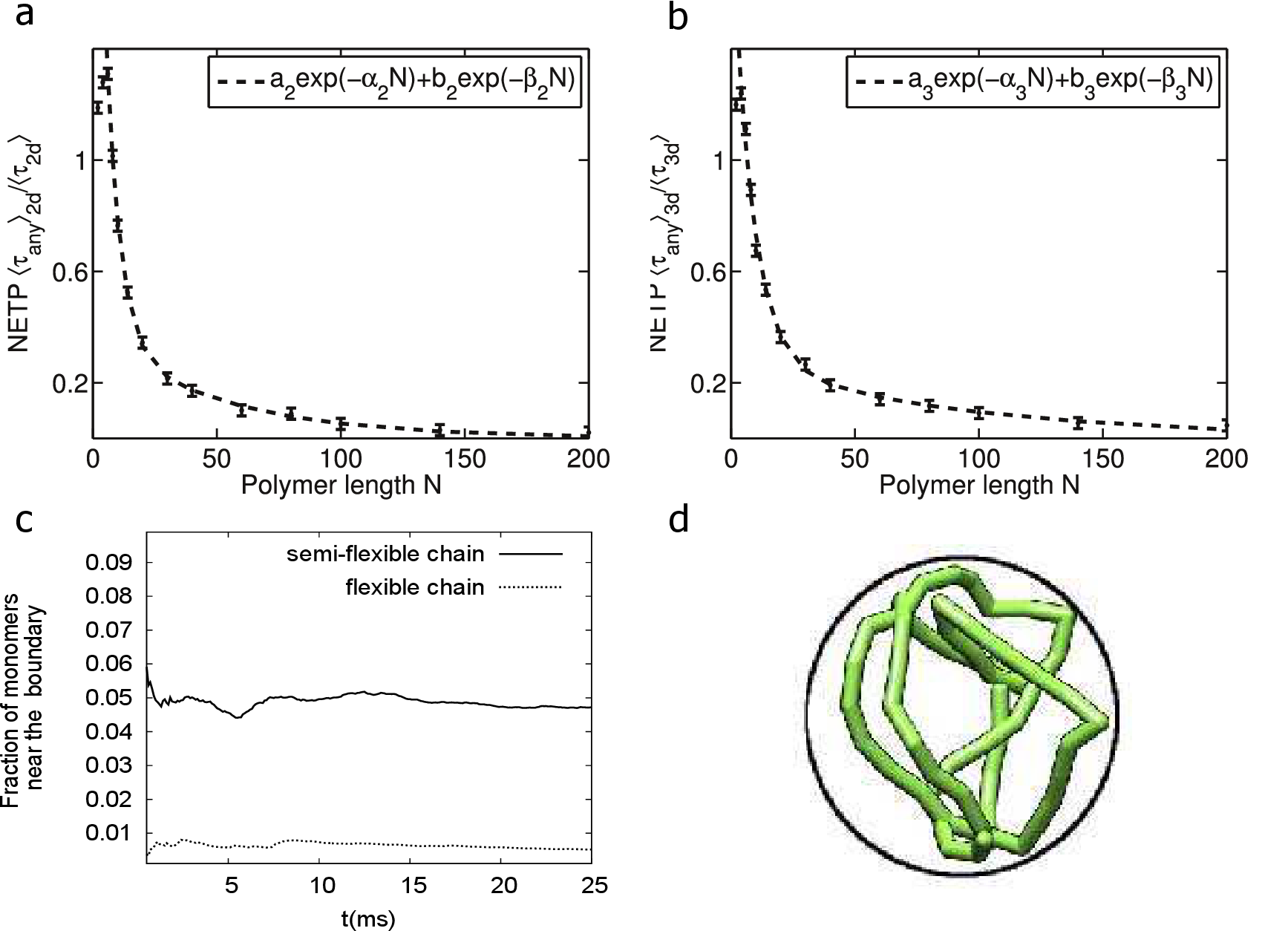}}
\caption{\textbf{NETP for a semi-flexible polymer}. The NETP is normalized to $\tau_0$ (the NET for 1 bead, equations \eqref{NET_d2} and \eqref{NET_d3}). The simulations are for any of the beads of a semiflexible polymer to reach the small target in dimension 2 (a) and 3 (b). Each data
point is an average over 2000 realizations. An double exponential fit leads to
$\frac{\langle \tilde \tau_2\rangle}{\langle \tau_{2d}\rangle} = a_2\exp(-{\alpha_2}N) +  b_2\exp(-\beta_2
N)$ with $a_2 = 0.37,b_2 = 2.9$ and $\alpha_2 = 0.02,  \beta_2 =
0.18$ and in dimension 3, $\frac{\langle \tilde \tau_3\rangle}{\langle \tau_{3d}\rangle} =
a_3\exp(-{\alpha_3}N) +  b_3\exp(-\beta_3 N)$. The same fit in three
dimensions gives $a_3 = 0.28, b_3 = 1.64$ and $\alpha_3 = 0.01, \beta_3 = 0.12$. {\bf (c)} Fraction of time a monomer spend near the boundary for a flexible and nonflexible polymer. {\bf (d)} snapshot of a nonflexible polymer, that has the tendency to locate near the spherical boundary.}
\label{any_stiff.fig}
\end{center}
\end{figure}

In this final section, we explore the consequence of adding flexibility on the NETP, which can characterize  complex DNA molecule containing various bound proteins. We account for the stiffness by including the bending energy
\beq
U_{bend}(\x) = \frac{\kappa_{ang}}{2}\sum_{i=1}^{N-1} (\mb{u}_{i+1} - \mb{u}_{i})^2 =     \kappa_{ang}\sum_{i=1}^{N-1}(1-\mb{u}_i \cdot \mb{u}_{i+1}),
\label{bending.eq}
\eeq
where $\mb{u_i}= \frac{ \x_{i+1} - \x_i }{|\x_{i+1} - \x_i|} $ is the
unit vector connecting two consecutive monomers and $\x_i$ is the
position of the $i-th$ monomer. This potential depends on the angle
$ \theta_i$ between two successive monomers with the relation
$\mb{u}_i \cdot \mb{u}_{i+1} = \cos\theta_i$.
The bending rigidity
$\kappa_{ang}$ is related to the persistence length $L_p$ of the polymer
by the expression \cite{Bohr}
\beq
L_p = \frac{\kappa_{ang}l_0}{k_BT},
\eeq
where $k_B$ is the Boltzmann constant, $T$ is the temperature, $l_0$ has been defined above.
The persistence length quantifies the stiffness of a polymer and
can be characterized using  the unit tangent vectors $t(s)$ and $t(0)$ at positions $s$ and $0$ along the polymer.  Averaging over all starting positions, the expectation of $\phi$, which is the angle between  $t(s)$ and $t(0)$, falls off exponentially with the distance $s$ along the polymer
\beq
\langle \hat{u}(s) \cdot  \hat{u}(0)\rangle =  e^{-s/L_P}.
\eeq
 For our simulations, we use the value $\kappa_{ang} = 5$ \cite{Stevens} that leads to $L_p = 250$nm, which is the reported value of the persistence length of chromatin fibers
\cite{ChromatinLp,ChromatinLp1}. We use the rigid polymer dynamics obtained from the over-damped
Langevin-Smoluchowski equation
\beq
\dot{\x} +\nabla U=\sqrt {2D} {\rm {\bf \dot {w}}},
\label{BD.eq_semi-flex}
\eeq
where the total potential $U = U(\x)$ is the sum of two energy potentials
\beq
U(\x) = U^N(\x) +  U_{bend}(\x),
\eeq
where $U^N(\x)$ is the elastic potential defined in
Eq.\ref{elastic}.
Interestingly, the two potentials are of the same
order of magnitude: Using the maximum extensibility of DNA about
10$\%$ of its total length \cite{Bustamante}, in that case the
length  $r-l_0 \sim  5$nm ($r = |x_{k+1}-x_{k}|$) and  $U_{el}  = \frac{1}{2}k (r - l_0)^2
\sim 2 \times 10^{-18}$Nm, while the maximum  energy between three
consecutive monomers due to bending  is  $U_{bend} =
k_{ang}(1-\cos\theta) = k_{B}T *L_p/l_0 * 2
\sim 5 \times 10^{-19}$Nm.
We simulate the arrival time of any monomer to the absorbing boundary and use the same fitting
procedure as in fig 2. It leads in dimension 2 to (Fig.\ref{any_stiff.fig}a)
\beq
\frac{\langle \tau_{\textrm{any}} \rangle_{2d}}{\langle \tau_{2d}\rangle} = a_2\exp(-{\alpha_2}N) +  b_2\exp(-\beta_2 N)
\eeq
with $a_2 = 0.37,b_2 = 2.9$ and $\alpha_2 = 0.02,\beta_2=0.18$
and in dimension 3,  (Fig.\ref{any_stiff.fig}b)
\beq
\frac{\langle \tau_{\textrm{any}} \rangle_{3d}}{\langle \tau_{3d}\rangle} = a_3\exp(-{\alpha_3}N) +  b_3\exp(-\beta_3 N)
\eeq
with $a_3 = 0.28, b_3 = 1.64$ and $\alpha_3 = 0.01,\beta_3 = 0.12$. Compared to the nonflexible polymer, the maximum of the NETP is now
shifted towards smaller values of $N$: The NETP is an increasing function of $N$ for $N < 6$, and for large $N$, it is a decreasing
function of $N$. We compare these results with the ones for flexible polymers: in a confined spherical cavity,
flexible polymers have a tendency to fill the available space, with a probability
of finding a monomer at the center of the cavity being higher than
at the boundary, when the persistence length is comparable to the
size of the sphere. On the contrary, for stiff polymers, the chain is forced to bend abruptly close to the surface and a large fraction of the polymer remains closed to the surface (Fig.\ref{any_stiff.fig}c-d).  Thus for a target located on the surface, a monomer located on a nonflexible polymer should reached it in an almost two-dimensional process, decreasing the NETP.

We conclude that large enough nonflexible polymer find small targets faster than completely flexible ones, this is due to an increase probability to find monomers in the close vicinity of the boundary where the small target is located.

\section*{Discussion and Conclusion}
We explored here by means of Brownian simulations, the mean time for part of a FJC polymer to find a small target located on the surface of a microdomain. {When each of the monomers can be absorbed,} our analysis reveals a bell-shaped curve (Fig. 2). For a short polymer, it must approach close to the target for one monomer to be absorbed and the main contribution to the NETP is given by the diffusion of the center of mass of the polymer $D_{\textrm{CM}}$. Since $D_{\textrm{CM}}$ is inversely proportional to $N$, the center of mass moves slower and the NETP increases with  $N$.   However, for longer polymers whose lengths are comparable to the size of the microdomain, some monomers can reach the target even if the center of mass is far away. In that regime, increasing the polymer length results in a {decrease} in the NETP.  It is surprising that the empirical fit of our Brownian simulations requires two exponentials. It would be interesting to derive analytically this law and the expression for the associated parameters. Another consequence of the present analysis is the approximation of the arrival time of a monomer to the target by a single exponential, showing the Poissonian nature of this process. This results is because hitting the target is rare event. We further showed that the polymer dynamics can drastically affect the NETP of a single monomer to the target. In that case, a repulsive force near the boundary is generated by the other reflecting monomers (Fig. 4). Consequently the NETP converges as the polymer increases to a constant value, which depends on the monomer position along the polymer chain.

Our analysis has several applications. In the context of chromatin dynamics in the nucleus, the long polymer regime shows that activation, defined as the arrival of one monomer to an active site by
the chromatin segment depends strongly on the chromatin length. Moreover, when the polymer represents a dsDNA break and the microdomain is the local confined domain generated around, our analysis shows that the mobile DNA segment (length of the polymer) significantly affects the search time: we found that increasing the length from small (free Brownian) to the maximum value leads to a factor larger that two difference(Fig. 2). To be quantitative, we can estimate the mean time for a  dsDNA break to find a specific target such as the other strand, with the following parameters: the free DNA (persistent) length is 50nm, it is located inside a chromatin sphere of radius 250nm, the diffusion coefficient is $D=4.10^{-2} \mathrm{\mu m}^2/\mathrm{s}$. We find using the  result of Fig 2 and formula \ref{emp1}-\ref{emp2} that the NETP (mean time for a monomer to find by diffusion a specific chromatin element of size 12.5nm or the other dsDNA break) is around $\tau=[100-200]$ seconds.

Finally, because a dsDNA break has been found to be directed to the nuclear boundary \cite{KarineDUBRANA}, we postulated that a specific mechanism should be involved. Indeed our analysis (Fig. 5) reveals that the reflective interaction of the polymer chain with the nuclear surface generates a local potential that prevents the break to approach the nuclear surface. Thus by increasing the surface of the target, which can results from dephosphorylation, leads to a decrease (Fig. 5) of the NETP of the break to the nuclear surface. It would interesting in a future study to add the effect of nucleosomes \cite{almouzni}. The present work disregards the effect of an external field (such as an electric field), which could be induced in vitro experiments. It would be interesting to add the effect of a field and explore the consequences on the NETP. In addition, the present study could also characterize the search of a target by a charged polymer. Indeed, because the Debye length is 0.8nm (about the size of a monomer), the polymer even if it charged cannot feel the target, showing that the present results are quite general.

\section*{Acknowledgement}
D. H. research is supported by an ERC starting grant.

\newpage
\section*{APPENDIX: NUMERICAL METHODS AND PARAMETER CALIBRATIONS}

\subsection*{Simulation procedure}
For a chain of $N$ beads (and $N-1$ springs), we initialized the
chain with its center of mass at the center of the domain, with the
angle between adjacent springs random (uniformly distributed between
0 and $2\pi$).

At each time step, the monomers (beads) move according to equation \eqref{eqnum}. when a Brownian bead crosses numerically the boundary of the domain, it is reflected according to Snell's law (angle of incidence $=$ angle of reflection). A monomer is "absorbed" at the
target region if the line drawn from its position at the previous
time step to its current position intersects with the target region
of the boundary. We record the first time any of the beads reaches the absorbing region as $\tau_\mathrm{any}$. If the bead is not the active monomer, it is reflected back into the domain and the simulation continues. If the bead is an active monomer, the simulation ends and the final simulation time is recorded as $\tau_\mathrm{mon}$. All our results are displayed in reduced units.
We have used the USFC Chimera software \cite{UCSF} to visual snapshots of the polymer dynamics.

\section*{Simulation parameters}

\subsubsection*{Dimensions of the microdomain:}
We use for the microdomain a disk and and a sphere in  dimension two and three respectively of radius $R = 250$nm.
For a size of the hole of 50nm, we obtain that $\varepsilon = \frac{a}{R}=0.05$.

\subsubsection*{Spring resting length $l_0$}
The resting length of the spring connecting two neighboring bead to be 50nm
(approximately 150 base pairs)\cite{DNA_persistence_length}.

\subsubsection*{Spring constant $k$}
Using the direct measurements \cite{DNA_spring_const} of the elasticity of short stretches (30 base pairs) of double stranded
DNA (dsDNA). The Young's Modulus of dsDNA is estimated to be 55 MPa \cite{DNA_spring_const}. We now estimate the spring constant $k$. Using that the dsDNA was on average initially 8.5nm, but when a force of 113.7pN
was applied, it stretches to an additional 6.5nm, under the assumption that the dsDNA is still in the Hookean
regime (i.e. stretching linearly) we obtain a spring constant of
 $k = F/∆x = 1.75 \times 10^{-2}$N/m.

\subsubsection*{Integration algorithm and time step $\Delta t$}
The simulation time step $\Delta t$ is chosen such that each bead moves on average
less than the distance $\Delta x^{\star}$ at each time step, where  $\Delta x^{\star}$ is the smaller
length scale of our system, which is the diameter of the small hole $2\varepsilon R$.
In practice, we used $\Delta x^{\star} = f\times (2\varepsilon R)$ where  $f$ is an extra precision parameter, fixed to 0.2. Hence we used $\Delta t  = (\Delta x^{\star})^2/2D = 1.5 \times 10^{-4}$s.
 The numerical scheme is as follows
\beq
{\bf \x}(t+\Delta t) = {\bf \x}(t)  +\frac{1}{\gamma}\nabla U^N \Delta t + \boldsymbol{\eta} \sqrt {2D \Delta t},
\eeq
that is (for one of the non-end beads),
\beq \label{eqnum}
{\bf \x}_{k}(t+\Delta t) =  {\bf \x}_{k}(t)-\frac{k}{\gamma} \left(  ({\bf \x}_{k}(t)-{\bf \x}_{k+1}(t))-l_0\frac{{\bf \x}_{k}(t)-{\bf \x}_{k+1}(t)}{|{\bf \x}_{k}(t)-{\bf \x}_{k+1}(t)|}  \right. \\
                                \left.   + ({\bf \x}_{k}(t)-{\bf \x}_{k-1}(t))-l_0\frac{{\bf \x}_{k}(t)-{\bf \x}_{k-1}(t)}{|{\bf \x}_{k}(t)-{\bf \x}_{k-1}(t)|} \right) \Delta t +    \boldsymbol{\eta} \sqrt{2D\Delta t} \nonumber
\eeq
where ${\bf \eta}$ is a Gaussian random variable with zero mean and unit variance.

\subsubsection*{The friction coefficient $\gamma$ and the diffusion constant $D$}

{The diffusion constant  $D$ of a DNA molecule has been estimated
\cite{DNAdiffusionconst} to be $D = 4\times 10^{-2} \, \mathrm{\mu}
\mathrm{m}^2/\mathrm{s}$. The friction coefficient is computed by
using Einstein relation $D = \frac{k_BT}{\gamma}$, at room temperature
and  $\gamma = 10^{-7}$Ns/m. The relaxation time
constant of each spring is $\ds{\frac{\gamma}{k}} \sim 10^{-5}$s,
much shorter than the time step we used.  Our simulations occur in a
regime where the springs  relax to their resting length $l_0$ at
every time step. Thus, the polymer keeps its original length
$(N-1)l_0$.}

\subsubsection*{Simulation of the stiff polymer in the microdomain}
We first introduced in the sphere a perfectly flexible polymer
($\kappa_{ang} = 0$), then we give it a stiffness by changing the
value of $\kappa_{ang}$ and a preliminary Brownian dynamics is
performed to ensure equilibration of the polymer.

\newpage
\begin{table*}
   \caption{Parameters of the simulations} \label{HApar}\vspace{0.5cm}
    \begin{tabular}{c|c|c|c}
    \textbf{Parameters} &  \textbf{Description}& \textbf{Value}\\
    \hline
      $R$ & Radius of the circular/spherical domain & $250\,\mathrm{nm}$ \\
      $a$ & radius of the absorbing window & $50\,\mathrm{nm}$  \\
      $l_0$  &Polymer persistence length & $50 \,\mathrm{nm}\;$ \cite{DNA_persistence_length} \\
      $D$ & Diffusion constant &  $4\times 10^{-2} \,  \mathrm{\mu m}^2/\mathrm{s}\;$ \cite{DNAdiffusionconst} \\
      $\gamma$ & Friction coefficient & $3.1 \times 10^{-5} \,\mathrm{Ns/m}$  \\
      $k$ & Spring constant& $1.75 \times 10^{-2} \ \mathrm{Nm}^{-1}\;$  \cite{DNA_spring_const} \\
\hline
      \end{tabular}
\end{table*}


\end{document}